\documentstyle[12pt]{article}

\begin{document}
\title{A Note on Non-commutativity}
\author{{\bf Valeri V. Dvoeglazov}\\
Universidad de Zacatecas\\
Apartado Postal 636, Suc. 3\\
Zacatecas 98061 Zacatecas, M\'exico\\
E-mail: valeri@fisica.uaz.edu.mx}

\date{\empty}
\maketitle


\begin{abstract}
ABSTRACT. Ambiguities have recently been found in the definition of the partial
derivative (in the case of presence of both explicit and implicit
dependencies of the function subjected to differentiation). 
We investigate the possible influence of this subject on quantum
mechanics and the classical/quantum field theory. Surprisingly, some
commutators of operators of space-time 4-coordinates and those of 4-momenta 
are {\it not} equal to zero.

RESUM\'E. Des ambigu\"it\'es ont \'et\'e r\'ecemment trouv\'ees dans la d\'efinition de la d\'eriv\'ee partielle (dans le cas de pr\'esence de d\'ependances \`a la fois explicites et implicites de la fonction soumise \`a la diff\'erenciation). Nous \'etudions l'influence possible de ce sujet sur la m\'ecanique quantique et la th\'eorie classique / quantique des champs. Fait surprenant, certains commutateurs 
d'op\'erateurs de coordonn\'ees spatio-temporelles et ceux de 4-moments ne sont pas \'egaux \`a z\'ero.


\end{abstract}

The assumption that the operators of
coordinates do {\it not} commute $[\hat{x}_{\mu },\hat{x}_{\nu }]\neq 0$
has been made by H. Snyder~\cite{snyder}. The Lorentz symmetry thus may be
broken. This idea~\cite
{noncom,kruglov} received attention in the context of ``brane theories''.
Moreover, the famous Feynman-Dyson proof of Maxwell equations~\cite{FD}
contains intrinsically the non-commutativity of velocities $[\dot x_i (t), 
\dot x_j (t)] \neq 0$ that also may be considered as a contradiction with
the well-accepted theories (while there is no any contradiction therein).

On the other hand, it was recently discovered that the concept of partial
derivative is {\it not} well defined in the case of both explicit and
implicit dependence of the corresponding function, which the derivatives act
upon~\cite{brown}.
The
well-known example of such a situation is the field of an accelerated
charge~ \cite{landau}.\footnote{%
Firstly, Landau and Lifshitz wrote that the functions depended on $t^{\prime }$,
and only through $t^{\prime }+R(t^{\prime })/c=t$ they depended implicitly
on $x,y,z,t$. However, later (in calculating the formula (63.7)) they used
the explicit dependence of $R$ on the space coordinates of the
observation point too. 
Jackson~\cite{jackson} agrees with~\cite{landau} that one should find ``a contribution to the
spatial partial derivative for fixed time $t$ from explicit spatial
coordinate dependence (of the observation point).''} 
\v{S}kovrlj and Ivezi\'{c}~\cite{ivezic} call this partial derivative as `{\it complete} partial
derivative'; Chubykalo and Vlayev,
as `{\it total} derivative with respect to
a given variable'.  The terminology suggested by Brownstein~\cite{brown} is
`the {\it whole}-partial derivative'.

Let us study the case when we deal with explicit and implicit dependencies 
$f ({\bf p}, E ({\bf p}))$. It is well known that the energy in the
relativism is connected with the 3-momentum as $E=\pm \sqrt{{\bf p}^2 +m^2}$%
; the unit system $c=\hbar=1$ is used. In other words, we must choose the
3-dimensional hyperboloid from the entire Minkowski space and the energy is 
{\it not} an independent quantity anymore. Let us calculate the commutator
of the whole derivative $\hat\partial /\hat\partial E$ and $\hat\partial / 
\hat\partial p_i$.
In order to make distinction between differentiating the explicit function
and that which contains both explicit and implicit dependencies, the `whole
partial derivative' may be denoted as $\hat\partial$.
In the general case one has 
\begin{equation}
{\frac{\hat\partial f ({\bf p}, E({\bf p})) }{\hat\partial p_i}} \equiv {%
\frac{\partial f ({\bf p}, E({\bf p})) }{\partial p_i}} + {\frac{\partial f (%
{\bf p}, E({\bf p})) }{\partial E}} {\frac{\partial E}{\partial p_i}}\, .
\end{equation}
Applying this rule, we surprisingly find 
\begin{eqnarray}
&&[{\frac{\hat\partial }{\hat\partial p_i}},{\frac{\hat\partial }{\hat%
\partial E}}] f ({\bf p},E ({\bf p})) = {\frac{\hat\partial }{\hat\partial
p_i}} {\frac{\partial f }{\partial E}} -{\frac{\partial }{\partial E}} ({%
\frac{\partial f}{\partial p_i}} +{\frac{\partial f}{\partial E}}{\frac{%
\partial E}{\partial p_i}}) =  \nonumber \\
&=& {\frac{\partial^2 f }{\partial E\partial p_i}} + {\frac{\partial^2 f}{%
\partial E^2}}{\frac{\partial E}{\partial p_i}} - {\frac{\partial^2 f }{%
\partial p_i \partial E}} - {\frac{\partial^2 f}{\partial E^2}}{\frac{%
\partial E}{\partial p_i}}- {\frac{\partial f }{\partial E}} {\frac{\partial%
}{\partial E}}({\frac{\partial E}{\partial p_i}})\,.  \label{com}
\end{eqnarray}
So, if $E=\pm \sqrt{m^2+{\bf p}^2}$ 
and one uses the generally-accepted 
representation form of $\partial E/\partial p_i
=  p_i/E$,
one has that the expression (\ref{com})
appears to be equal to $(p_i/E^2) {\frac{\partial f({\bf p}, E ({\bf p}))}{%
\partial E}}$. Within the choice of the normalization the coefficient may be related to 
the longitudinal electric field in the helicity basis (the electric/magnetic
fields can be derived from the 4-potentials which have been presented in~ 
\cite{hb}). On the other hand, the commutator 
\begin{equation}
[{\frac{\hat\partial}{\hat\partial p_i}}, {\frac{\hat\partial}{\hat\partial
p_j}}] f ({\bf p},E ({\bf p})) = {\frac{1}{\vert E\vert^3}} {\frac{%
\partial f({\bf p}, E ({\bf p}))}{\partial E}} [p_i, p_j]\,.
\end{equation}
This should also not  be zero according to
Feynman and Dyson~\cite{FD}. They postulated that the velocity (or, of course, the 3-momentum)
commutator is equal to $[p_i,p_j]\sim i\hbar\epsilon_{ijk} B^k$, i.e., to
the magnetic field.
In fact, if we put in the corespondence to the momenta their
quantum-mechanical operators (of course, with the appropriate clarification $%
\partial \rightarrow \hat\partial$), we obtain again that, in general, the
derivatives do {\it not} commute $[{\frac{\hat\partial}{\hat\partial x_\mu}}%
, {\frac{\hat\partial}{\hat\partial x_\nu}}] \neq 0$.

Furthermore, since the energy derivative corresponds to the operator of time
and the $i$-component momentum derivative, to $\hat x_i$, we put forward the
following Ansatz in the momentum representation: 
\begin{equation}
[\hat x^\mu, \hat x^\nu] = \omega ({\bf p}, E({\bf p})) \,
F^{\mu\nu}_{\vert\vert}{\frac{\partial }{\partial E}}\,,
\end{equation}
with some weight function $\omega$ being different for different choices of
the antisymmetric tensor spin basis. In the modern literature the relation 
$[x^\mu, x^\nu]\sim F^{\mu\nu}$, the electromagnetic tensor, is frequently 
used~\cite{elmnoncom}.
However, the idea of the broken Lorentz invariance by this
method~\cite{prl} concurs with the idea of the {\it fundamental length}, first
introduced by V. G. Kadyshevsky~\cite{kadysh} on the basis of old papers by
M. Markov. Both ideas and corresponding theories are extensively discussed,
e.g.~\cite{amelino}. In my opinion, the main question is: what is the space
scale, when the relativity theory becomes incorrect.

\section*{Conclusions}

We found that the commutator of two derivatives may be {\it not} equal to
zero. As a consequence, for instance, the question arises, if the derivative 
$\hat\partial^2 f/\hat\partial p^\nu\hat\partial p^\mu$ is equal to the
derivative $\hat\partial^2 f/\hat\partial p^\mu\hat\partial p^\nu$ in all
cases?\footnote{%
The same question can be put forward when we have differentiation with
respect to the coordinates too, that may have impact on the correct
calculations of the problem of accelerated charge in classical
electrodynamics.} The presented consideration permits us to provide some
bases for non-commutative field theories and induces us to look for further
development of the classical analysis in order to provide a rigorous
mathematical basis for operations with functions which have both explicit
and implicit dependencies.

I am grateful to participants of conferences where this idea has been discussed~\cite{dvoeglazovaps}.
I greatly apreciate the referee clear reports.
I am grateful to R. Keys for his help with the French translation.

\end{thebibliography}
\end{document}